\documentclass[11pt,twoside]{article}
\usepackage{asp2004}
\usepackage{psfig}
\usepackage{epsf}
\usepackage{graphics}
\usepackage{lscape}
\def\HI{H{\,\small I}}

\newcommand{\msun}{{$M_\odot$}}
\newcommand{\kms}{$\,$km$\,$s$^{-1}$}
\newcommand{\ltsima} {$\; \buildrel < \over \sim \;$}
\newcommand{\gtsima} {$\; \buildrel > \over \sim \;$}
\newcommand{\lta} {\lower.5ex\hbox{\ltsima}}
\newcommand{\gta} {\lower.5ex\hbox{\gtsima}}

\def\edcomment#1{\iffalse\marginpar{\raggedright\sl#1\/}\else\relax\fi}
\reversemarginpar

\begin{document}
\title{
Outflows of neutral (and ionized) gas in radio galaxies}
\author{R. Morganti, T. Oosterloo}
\affil{Netherlands Foundation for Research in Astronomy, Postbus 2,
NL-7990 AA, Dwingeloo, The Netherlands}
%{\tt morganti,oosterloo@astron.nl}}
\author{C.N.  Tadhunter}
\affil{Dep. Physics and Astronomy,
University of Sheffield, S7 3RH, UK}

\begin{abstract}

Outflows  up to 1500 \kms\ of atomic neutral hydrogen are
detected in a growing number of radio galaxies. Outflows with similar
velocities are also detected in ionized gas, suggesting a common origin
for the extreme kinematics of these two phases of the gas.  The high detection
rate of
such outflows  in young (or restarted) radio sources appears to be related 
to the existence of a dense ISM around these objects.  Such a dense ISM can
have important consequences for the evolution of the radio source and
the galaxy as a whole.  Here we summarize the recent results obtained
and the characteristics derived so far for these outflows.  We also
discuss possible mechanisms (e.g. interaction between the radio plasma
and the ISM and adiabatically expanding broad emission lines clouds)
that can be at the origin of these phenomena. 

\end{abstract}
\thispagestyle{plain}

\section{Introduction}

The presence of neutral hydrogen in the central regions of active
galaxies is often explained as being due to gas in circumnuclear tori/disks,
i.e. with gas relatively settled and regularly rotating (see
e.g. Pihlstr\"om, Conway, Vermeulen 2003, Mundell et al. 2003 and
Morganti et al. 2004a for a review).  However, the conditions of the
\HI\ are often more complex and disturbed  kinematics of the
atomic neutral hydrogen is observed in a growing number of radio
sources.  The presence of this ``unsettled'' gas can now be detected
thanks to the possibility of performing, with sensitive systems,
\HI-21 cm observations using broad spectral bands (e.g. bands that can
cover up to $\sim 4000$ \kms\ in the case of the recently upgraded WSRT).

The presence of gas with extreme kinematics in the central regions of
active galactic nuclei (AGN) has recently attracted particular
attention, in particular on the presence and occurrence of fast gas
outflows.  Fast nuclear outflows of {\sl ionized} gas appear to be a
relatively common phenomena in active galactic nuclei (see
e.g. Crenshaw et al.\ 2000, Kriss et al. 2004, Capetti et al. 1999,
Krongold et al.\ 2003, Veilleux et al.  2002, Tadhunter et al. 2001
Elvis 2000 and ref. therein).  They are mainly observed in optical, UV
and X-ray observations.  Gas outflows associated with AGN provide
energy feedback into the interstellar medium (ISM) that can profoundly
affect the evolution of the central engine as well as that of the host
galaxy. The mass-loss rate from these outflows can be a substantial
fraction of the accretion rate needed to power the AGN.  Thus, the
physics of these phenomena needs to be understood in order to
understand AGN.
It is not too surprising to find such outflows also in radio galaxies
(see Tadhunter et al. 2001, Holt et al. 2003 and Morganti et al. 2003a
for a summary of recent results).  However, it is extremely intriguing
that is several radio sources fast outflows of {\sl neutral} hydrogen
( up to 2000 \kms) have been discovered.  The physical conditions of
this gas provide new and important clues on the mechanism responsible
for AGN-related outflows.
Here, we summarize the cases found so far and the characteristics of
the neutral hydrogen associated with such outflows. We also discuss
some of the possible mechanisms that can be at the origin of this
phenomenon.

\begin{figure*}
\centerline{\psfig{figure=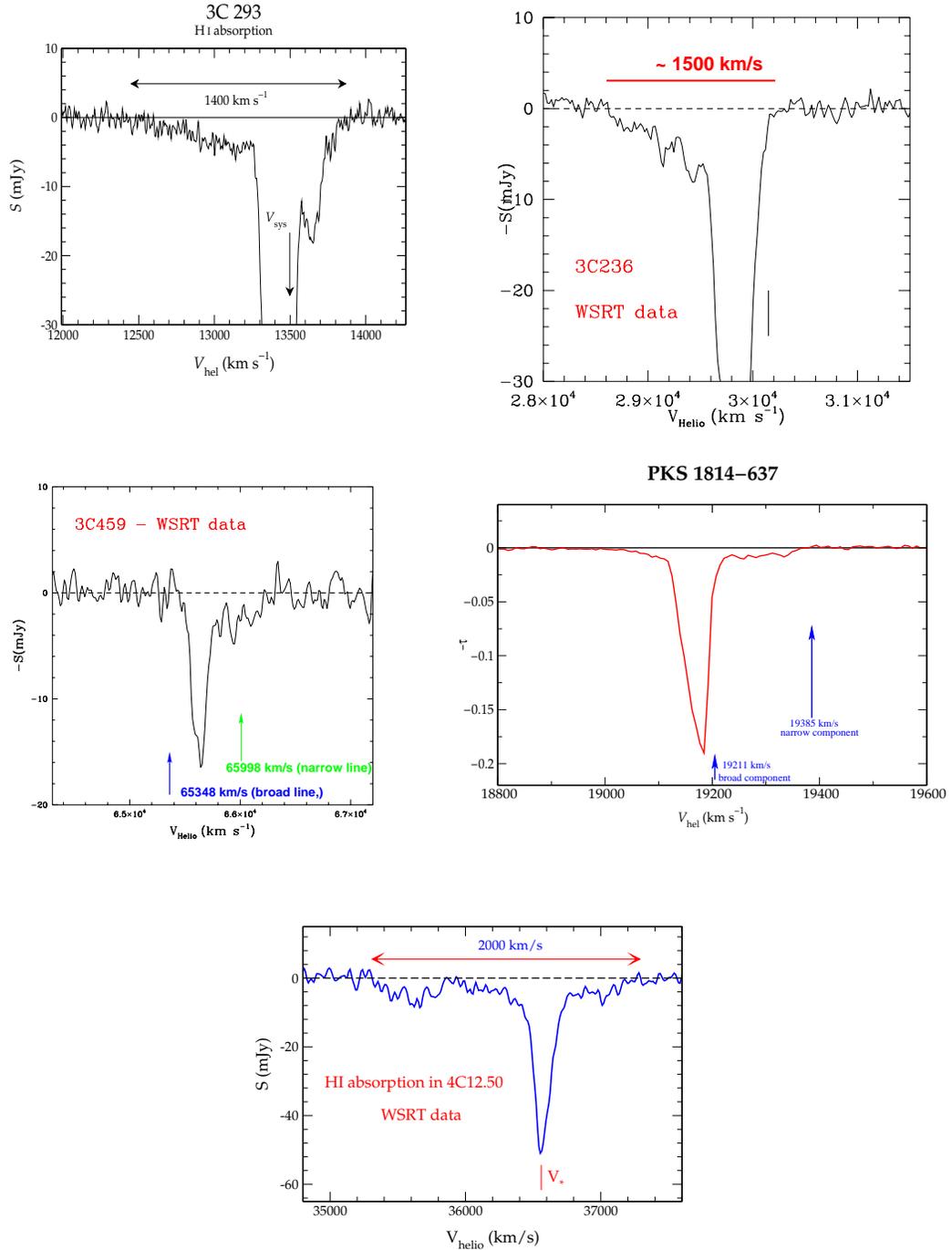,angle=0,width=16.0cm}
}
\caption{Examples of broad \HI\ absorption has been
detected in radio galaxies.  The systemic velocity as derived from optical
emission lines is also indicated. In two cases (3C~459 and PKS~1814-637) we
have marked both the systemic velocity (corresponding to the velocity of the
{\sl narrow} lines) as well as the velocity of the {\sl broad} component
also detected in the optical emission lines (Holt et al. in prep.). See text for details.}
\end{figure*}

\section{Characteristics and occurrence of the broad \HI\ absorption}

The first case of \HI\ outflow has been found in a radio loud Seyfert
galaxy, IC~5063 (Morganti et al. 1998). The identification of the
broad \HI\ absorption with an outflow has been possible because of the
accurate value of the systemic velocity available for this galaxy.
This object is one of the best studied so far. It has been followed up
with the Australian LBA and these high resolution observations have
confirmed that the location of the absorption is against the brighter
(eastern) radio lobe, few hundred pc from the nucleus (Oosterloo et
al. 2000) . More recently, an optical spectroscopic study (using the
NTT) has shown that, at the same location, also the ionized gas has
very complex kinematics indicating an outflow of similar velocity as
for the neutral gas (Morganti et al. 2003b).

In radio galaxies, we have detected so far 8 cases of broad (up to
2000 \kms) \HI\ absorption.  In all but one of these objects (see
Sec. 2.1), most of the absorption appears blueshifted compared to the
systemic velocity and therefore associated with outflowing gas.  The
outflow velocities are up to 1500 km/s. The column densities of these
broad absorption are very low, typically $\tau \sim 0.001 -
0.005$. With the sensitivity of present-day radio telescopes, this
limits the detection of such absorption to very strong radio sources.
These broad \HI\ absorption features have been mainly detected using
the broad band (20~MHz) receiver now available at the WSRT. However,
given the limited spatial resolution of these observations ($\sim 10$
arcsec), the exact locations of the absorption is not known.  Indeed,
most of these sources have a relatively complex continuum structure on
the sub-arcsec and sometimes milliarcsec scale.  In the case of the
nearby radio galaxy 3C~293, described in detail in Emonts et al. (this
Volume), the \HI\ absorption could be located as far as $\sim 1$ kpc
from the nucleus, similar to what found in the Seyfert galaxy IC~5063.
However, there are also cases where the broad \HI\ absorption can be
located much closer to the nucleus.  An interesting example is the
Compton-thick, broad-line and GigaHertz Peaked (GPS) radio galaxy
OQ~208. This source is only 10 pc in size (Stangellini et al. 1997)
and the \HI\ spectrum is shown in Fig.~2.  More cases where the broad
\HI\ absorption occurs close to the nucleus may exist. However, high
resolution follow-up observations are needed to investigate this.

The lack of detailed information about the location of the broad \HI\
absorption implies an uncertainty in deriving the physical parameters
of the gas outflows. Assuming that the absorption uniformly covers the
radio source and a T$_{spin} = 100$ K, the column density associated
with the absorption is typically around few times 10$^{20}$
cm$^{-2}$. These values can easily go up to few times 10$^{21}$ --
10$^{22}$ cm$^{-2}$ if the absorption is instead localized in a small
area.  Moreover, the T$_{spin}$ can have much higher values (1000 K or
more) if the absorption occurs very close to the nucleus.  The density
of the \HI\ (still very dependent on the location and size of the
absorption) ranges from $\sim 0.2$ cm$^{-3}$ (this could be the case
in 3C~293) to $\sim 30$ cm$^{-3}$ (in the case of OQ~208).  The \HI\
mass associated with the outflows ranges from $\sim 10^3$ up to $10^6$
\msun.  The energy flux associated with the \HI\ outflows is of the
order of $10^{40} - 10^{41}$ erg/s.

It is interesting to notice that, although in the cases described above (and
shown in Figs. 1 and 2) most of the broad \HI\ absorption is associated with
blueshifted gas (indicating an outflow), indication of redshifted wings are
also seen in some objects although with much smaller amplitude.  One exception to
this is the case of the compact radio galaxy 4C~37.11 where the new WSRT
observations show a broad \HI\ absorption of about $\sim 1500$ \kms\ (see
Fig.~3) symmetrical around the systemic velocity. Interestingly, VLBA
observations (Maness et al. 2003) recover only part of the \HI\ absorption
detected by the low resolution WSRT spectrum.

\begin{figure*}
\centerline{
\psfig{figure=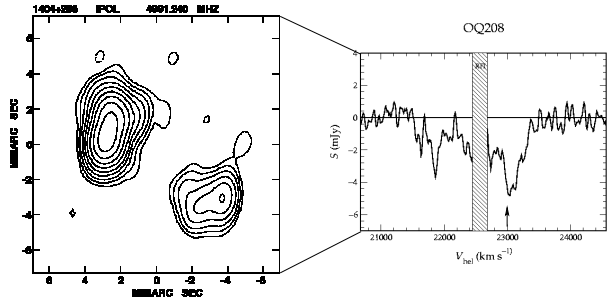,angle=0,width=8cm}}
\caption{
Broad \HI\ absorption detected (using the WSRT) against the compact
source OQ~208. The systemic velocity derived by Marziani et al. (1997) is
also indicated. The VLBI radio continuum (that is only $\sim 10$ pc in
size) is taken from Stanghellini et al. (1997).}
\end{figure*}

\begin{figure*}
\centerline{
\psfig{figure=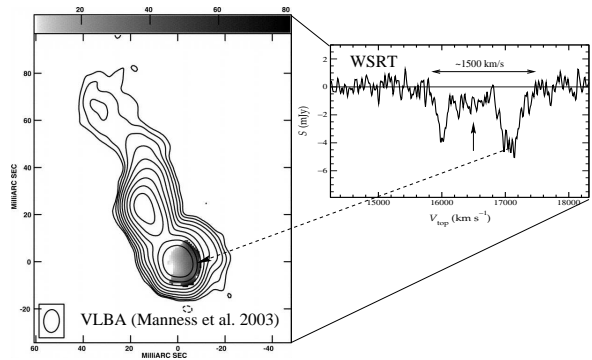,angle=0,width=8cm}}
\caption{1.3 GHz continuum map overlaid by the \HI\ opacity distribution over 
the source as 4C~37.11 detected by the VLBA (from Maness et al. 2003).
The absorption correspond to the most redshifted component seen in the WSRT
profile shown in the figure on the right. }
\end{figure*}

\section{Combining neutral and ionized gas}

As mentioned above, direct evidence for fast outflows has been found
also in the near-nuclear \emph{ionized} gas from the detailed analysis
of the strong optical emission lines characteristic of powerful radio
galaxies. These studies further show how complex are the conditions of
the gas in the circum-nuclear regions by revealing the presence of
high ionization [OIII] emission lines with very broad (1000 to 2000
km/s) and blueshifted by 400- 2000 km/s relative to the extended,
quiescent gaseous halos (Tadhunter et al. 2001, Holt et al. 2003).
The presence of more than one kinematic component of the ionized gas
(a narrow and a broad component detected in the emission lines) is
indicated for two objects in Fig.~1 but more cases have been found
(see below).  The difference in velocities between these two
components can be substantial and it is therefore important to derive
the systemic velocity from the component associated with the more
quiescent (i.e. narrower line profile) gas.  The accurate value of the
systemic velocity is, of course, crucial for the interpretation of the
kinematics of the gas (both ionized and neutral). It is, therefore,
clear how important these combined studies of the neutral and ionized
gas are for a comprehensive interpretation of gas conditions around
AGN.  Accurate values of the systemic velocity are now available for all
the galaxies for which broad \HI\ absorption has been found.

The most extreme case of complex kinematics of the ionized gas has been
observed in the compact radio galaxy 4C~12.50 (see Holt et al. 2003
for details). In this object, complex optical emission line profiles
-- with at least three kinematical components, as broad as 2000 \kms\
and blueshifted $\sim 2000$ \kms\ -- have been found at the position
of the nucleus. These components have been identified - through
measurement of density, reddening and the association with the neutral
hydrogen components - as coming from different regions. These regions
range from a large-scale quiescent halo to a inner region strongly
affected by the radio jet (and associated with the most kinematically
disturbed component). All this supports the idea that 4C~12.50 is a
young radio source with the nuclear region still enshrouded in a dense
cocoon that is in the process of being swept away (see also
Sec.~5). We may be looking, therefore, at the first stage of the
evolution of the newly born (or restarted) radio jet that will evolve
into a Cygnus A-like object (Tadhunter et al. 1999) where the result
of the outflows in the form of hollowed out biconical structures can
be seen.

Interestingly in 4C~12.50 some similarities can be seen between the
kinematics of the neutral and ionized gas. Another similar case is
3C~293 (see Emonts et al. this Volume).  The similarities between the
two phases of the gas may indicate that both outflows are originating
from the same mechanism and could be co-spatial. If confirmed for
other galaxies it would provide further constrains for the models that
should describe these phenomena.

\begin{figure*}
\centerline{\psfig{figure=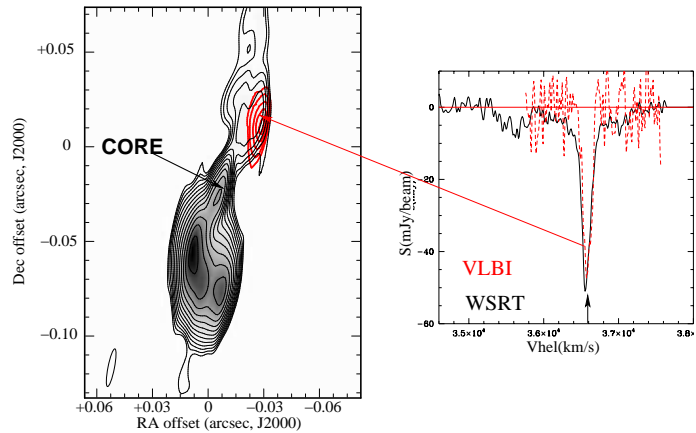,angle=0,width=10cm}}
\caption{{\sl Left} VLBI continuum image (grey scale and thin contours) of
4C~12.50  superimposed onto the total
intensity of the (narrow) \HI\ absorption (thick contours). See  Morganti et al. (2004b) for more details. The position of the radio core is also
indicated.  {\sl Right} \HI\ absorption
profile observed with the WSRT (black) and VLBI (red). A broad,
shallow \HI\ absorption is detected in the WSRT observations (see also Fig.~1). Due to
the narrower band, this broad absorption is not detected in the VLBI
observations.
}
\end{figure*}

\section{Origin of the \HI\ outflows}

The central question is how {\sl neutral} gas can be associated with
such fast outflows. As discussed in Morganti et al.\ (2003c) for the
case of 3C~293, a number of possible hypotheses can be made. They
include: starburst winds (Veilleux et al. 2003), radiation
pressure from the AGN (Dopita et al.\ 2002), jet-driven outflow and
adiabatically expanding broad emission line clouds (Elvis et al.\
2002).  For the cases where the broad absorption seem to happen far
away (few hundred pc) from the nucleus (e.g. 3C~293), the model that
we favour is jet-driven outflow. The energy flux associated with the HI
outflows (although uncertain) can be easily accounted for by the the
energy carried by the radio plasma, that in the case of the
radiogalaxies studied so far it is at least of the order of $10^{42}$
erg/s. However, regardless the energetics, the \textsl{neutral} gas
needs to be accelerated to velocities many times its local sound
speed: how this is done is not yet clear. A possibility is that the
radio plasma jet hits a (molecular) cloud in the ISM.  As a
consequence of this interaction, the kinematics of the gas are
disturbed by the shocks.  Once the shock has passed, part of the gas
may have the chance to recombine and become neutral while it is moving
at high velocities. The possibility that cool gas clouds will be
formed in this way has been explored via numerical simulations with the
main aim of investigating whether star formation can be produced via
jet-cloud interaction (see Mellema et al.\ 2002 and Fragile et
al. 2004).  Although it seems to be possible to form cool gas (and
\HI) this way, it is still unclear whether it is also possible to
accelerate the gas clouds to the high velocities that we observe.

However, in other cases (like e.g. OQ~208) the outflows appears to
happen on much smaller scales. In these cases the outflows could
originate from radiation pressure for example from adiabatically
expanding broad line clouds.  Elvis et al. (2002) have investigated the
evolution of such clouds and they derive these cloud would reach
temperatures of $\sim 1000$ K at distances of the $\sim 3$ pc from the
nucleus. When they reach that phase they would be able to form
dust. If the clouds can expand further, they will also cool enough to
produce \HI.

While for the jet/cloud interaction we may expect to find the broad
\HI\ absorption in the vicinity of strong radio continuum structure,
in the case of radiation pressure this does not have to be the case
and the accelerated gas may be found mainly close to the
nucleus. These scenarios can be further tested with high-resolution
observations.  Whatever mechanism is responsible to produce the
outflow of atomic hydrogen, the presence of such a component will put
some tight constrain on the models that have to describe such a
phenomenon.

\section{Relevance for the evolution of the radio sources}

A particularly high detection rate of broad \HI\ absorption has been
found in radio galaxies considered to be either in the early-stage of
their evolution (like 4C~12.50) or in a phase of re-started activity
(perhaps the case for 3C~293). This is suggested, e.g., by the
detection in these objects of a ``young'' ($\sim 1$ Gyr old) stellar
population (Tadhunter et al. 2004). These objects appear to have a
particularly rich medium -- they are often detected in CO and are
typically far-IR bright, Evans et al. (2004) -- likely resulting from
a (major) merger that happened in their recent past. The \HI\ can be a
further tracer of this rich medium and the complex kinematics of the
\HI\ and ionized gas results from the interaction between the energy
released by the AGN and the dense ISM.

The case of 4C~12.50 is again particularly interesting. This radio
galaxy has often been suggested to be a prime candidate for the link
between ultraluminous infrared galaxies and young radio galaxies. In
this object, even the deep and relatively narrow \HI\ absorption
(observed at the systemic velocity) has been found to be associated
with an off-nuclear cloud ($\sim 50$ to 100 pc from the radio core)
with a column density of $\sim 10^{22}\ T_{\rm spin}/(100\ {\rm K}$)
cm$^{-2}$ and an \HI\ mass of a few times $10^5$ to $10^6$ \msun\ (see
Fig.~4 and Morganti et al. 2004b). There are more examples of objects
where the \HI\ traces the rich medium surrounding the active
nucleus. Examples of off-nuclear \HI\ absorption are found in 3C~236
(Conway \& Schilizzi 2000) and, more recently, against the southern
hot-spot of the CSO 4C~37.11 (Maness et al.\ 2004, see also Fig.~3).
This may have important implications for the evolution of the radio
jets. Although this gas will not be able to confine the radio source,
it may be able to momentarily destroy the path of the jet as shown
also by numerical simulations (Bicknell et al.\ 2003). Thus, this
interaction can influence the growth of the radio source until the
radio plasma clears its way out.
A similar situation may occur in the case of OQ~208. Guainazzi et al. (2004)
suggest that in this source we could be seeing the jets piercing their way
through a Compton-thick medium pervading the nuclear environment.  The outflow
detected in \HI\ (see Fig.~2) would be an other indication of this process.
Guainazzi et al. (2004) also suggest that if the jets have to interact with
such a dense medium, one could largely underestimating the radio activity
dynamical age determinate for this kind of sources from the observed hot-spot
recession velocity.

\end{document}